\documentclass[aps,preprint,showpacs,superscriptaddress,groupedaddress]{revtex4}  % for double-spaced preprint
\usepackage{graphicx}  % needed for figures
\usepackage{dcolumn}   % needed for some tables
\usepackage{bm}        % for math
\usepackage{amssymb}   % for math
\begin{document}

\title{Change of surface critical current in the surface superconductivity and mixed states of superconducting Niobium.}
\author{Muhamad Aburas, Alain Pautrat}
\email{alain.pautrat@ensicaen.fr}
\thanks{Corresponding author}
\author{Natalia Bellido}
\affiliation{Laboratoire CRISMAT, UMR 6508 du CNRS, ENSICAEN et Universit\'{e} de Caen, 6 Bd Mar\'{e}chal Juin, F-14050 Caen 4.}

\begin{abstract}
A systematic study of irreversible magnetization was performed in bulk Niobium after different surface treatments.
Starting with smooth surfaces and abrading them, a strong increase of the critical current is observed up an apparent limiting value.
 An impressive change of the critical current is also observed in the surface superconductivity (SSC) state, reaching values of the same order of magnitude as in the mixed state.
 We explain also the observation of strong SSC for magnetic field perpendicular to larges facets in terms of nucleation of SC along bumps of a corrugated surface.
 
\end{abstract}

\newpage
\maketitle 

\section{Introduction}

 Vortex pinning in superconductors of the second kind has been deeply investigated by the scientific community.
The vortex lattice is a disordered elastic model system with several parameters which can be tuned by the experimentalist \cite{shobo}. Investigating the vortex physics has also very practical issues, since vortex pinning controls the critical current, the linear ac response and all the non linear transport properties of superconductors of the second kind \cite{EH}.
 There are several defects which can act as potential pinning centers. In the literature, bulk defects are the most often considered, specially in the high Tc materials where a large number of bulk defects and/or composition fluctuations can be present in non stoichiometric oxides \cite{Kees,Kes}. The important role of sample surfaces and boundaries have been also emphasized \cite{YBCOsurface1,YBCOsurface2,Flippen}. This effective role of samples boundaries, in particular the surface quality, has been enlightened in low Tc materials some years ago \cite{LowTcsurf}, but more rarely addressed since then \cite{Niobiumnous}. The case of Niobium (Nb) is particularly relevant for studying the effect of surfaces for the critical current since Nb is well known for its use in radiofrequency cavities, where the role of surface quality for optimized performances is specially critical \cite{Halbritter}. The possibility of Niobium to sustain large subcritical currents for  magnetic fields above Bc2, i.e. in the surface superconducting (SSC) state, deserves also a special attention \cite{casalbuoni}. SSC can be also the place of very peculiar and interesting properties as reported in \cite{pan,relax,niobsurfnoise}.
 
  In this paper, we investigate the consequence of simple surface treatments for the critical current and vortex pinning in a bulk Nb sample, extending the measurements also in the SSC state where the bulk sample goes into its metallic state. In particular, evidence that SSC currents can be impressively tuned by surface roughness is reported. This shows also that surface roughness is the cause of unexpected SSC in the geometry where B is perpendicular to the large facets.

 \section{Experimental}

 All the samples under study are cut from the same ingot of high purity polycristalline bulk Niobium previously used for transport and noise measurements \cite{Niobiumnoise} or small angle neutron scattering \cite{TPA}. 
 In order to start with samples of similar surface states and with a moderate roughness, a buffer chemical
polishing (BCP) was first performed on all the samples. It consists in plugging them into a solution of $H_3PO_4 (85\%), HF (40\%), HNO_3 (69\%)$
for eight minutes. After this BCP, the surfaces exhibit a typical RMS roughness close to 1 $nm$. We measure 0.7 $nm$ RMS over 10*10 $\mu m^2$ using AFM technique.

Our Niobium thermodynamical parameters (T$_c$, B$_{c2}$, B$_{c3}$) were then measured on the same sample (sample $\sharp$1) using different techniques: DC resistivity, V(I) curves, magnetization (SQUID) and specific heat measurements. After these measurements, different surface treatments have been performed on 3 samples. Note that the surface treatments have been performed only on the large surfaces, and not on the small lateral surfaces. To compare their critical currents, we have measured magnetization loops used a SQUID magnetometer. For most of the measurements, B is perpendicular to these large surfaces (B$_{perp}$ geometry), otherwise it is be specifically written in the text (B$_{para}$ geometry for B parallel to the large facets in the fig.5). Calculation of the critical current density J$_c$ was made using the irreversible part of the volume magnetization $\Delta M$ with the modified Bean model for a slab geometry \cite{EH}. For a sample surface a $\times$ b (a$<$b) and a thickness t, J$_c$ can be approximate by:
\begin{equation}
J_c \approx 20/(4\pi 10^{-4}) \Delta M / (a-a^2/3b)
\end{equation} 
with J$_c$ in A/m$^2$, $\Delta$M in A/m and a, b in m.

\section{Results and discussion}
\subsection {Characterization and critical current}
In Fig.1 are shown different measurements performed to identify both the second critical field B$_{c2}$ and the critical field of surface superconductivity B$_{c3}$.
From the specific heat, B$_{c2}$ is clearly identified as the onset of bulk superconductivity. However, under a low applied current, the electric resistance reaches its normal state value R$_n$ at much higher field, and hysteric magnetization is also observed up to B$_{c3}\approx$1.7 B$_{c2}$ due to the existence of a finite critical current in the surface superconducting (SSC) state \cite{Rollins,Hart,patrice}. From V(I) curves (not shown here), we find that the differential resistance dV/d(I-I$_c$)$\approx$ R$_n$ as soon as B$>$B$_{c2}$ as expected for a metallic bulk \cite{niobsurfnoise,kim}. Then, from these different measurements, it is possible to identify both critical fields B$_{c2}$ and B$_{c3}$ at different temperatures.  This allows to draw the phase diagram of a Niobium sample $\sharp$1 in Fig.2. If the hysteretic magnetization allows to extract the values of persistent currents in the SSC, we note that a more precise value of B$_{C3}$ could be obtained using ac susceptibility \cite{acsusc}. If we compare our B$_{c3}$/B$_{c2}$ $\approx$1.70 with B$_{c3}$/B$_{c2}$ $\approx$1.86 reported by S. Casalbuoni et al \cite{casalbuoni} on BCP Niobium, we find a value much closer to the Ginzburg Landau limit of Saint James-De Gennes. Assuming that the change of the change of B$_{c3}$/B$_{c2}$ is due impurities concentration close to the surface as in \cite{casalbuoni}, one possible explanation is that the rather short duration of the BCP that we used ($\approx$ 8 min) limits the impurities over a depth less than 1 nm, much smaller than the coherence lenght.  

In the SSC state, superconductivity nucleates along a surface sheath over a typical scale of 0.7$\xi$ \cite{degennes}. In the present geometry, B is perpendicular to the large facets of the sample, whereas SSC is theoretically expected for B parallel to the surfaces. This feature has been already reported in Nb \cite{niobsurfnoise}, NbSe$_2$ \cite{danna}, MgB$_2$ \cite{MgB2} but the explanation is not straightforward. This was tentatively attributed to surface roughness over a scale of $\xi$ along the perpendicular facets \cite{niobsurfnoise} or to inhomogeneous path of current along the lateral facets \cite{danna}. 

After a BCP, 3 samples were selected for different grinding and "polishing" processes (table 1). We have used 2400 sandpaper, 3 $\mu$m diamond paste and colloidal silica with particle sizes of nominally 50 nm in diameter. For the colloidal silica process, a progressive abrasion was first made with a rough polishing paper (4000) and then with alumina (1 $\mu$m followed by 0.3 $\mu$m), then the final polishing was made by hand during 5 minutes.  Then the irreversible magnetizations were measured to compare their critical current J$_c$  using the extended Bean model. Note that all the samples having the same dimensions within few percent after the surface treatments, the relative variation of critical current between all the samples is very similar to that of the irreversible magnetization. To have samples with similar thicknesses, and because the colloidal silica polishing removes a significant part (typically 150 $\mu$m), we started with a thicker slab for the sample $\sharp$4. 

\begin{table}[h]
\begin{center}
\begin{tabular} { | c | c | }
\hline
Surface treatment & sample: width$\times$length$\times$thickness (mm)\\
\hline
chemical polishing & $\sharp$1:  4.26$\times$5.42$\times$0.28\\
\hline
 2400 sandpaper & $\sharp$2:  4.67$\times$4.85$\times$0.22\\
\hline
 3 $\mu$m diamond & $\sharp$3:  4.34$\times$4.98$\times$0.21 \\
\hline
colloidal silica & $\sharp$4:  4.40$\times$4.48$\times$0.21\\
\hline
\end{tabular}
\end{center}
\caption{\label{tab:1} surface treatments (see text for details) and corresponding sample reference and size.}
\end{table}

From the J$_{c}$ value, and since only the surfaces of the samples were altered, we express the critical current as a superficial critical current per unit of width $i_c$=J$_c$ $\times$ t/2 with t the thickness of samples.
We report in Fig.3 the variation of $i_c$ as function of the applied magnetic field at T= 6K for all the samples. Note that the relative variation between the critical current of all the samples is similar at 2K and 4K , but the irreversible magnetization in the Shubnikov state (B$<$B$_{c2}$) is affected by strong flux jumps at 4K and 2K especially for samples $\sharp$2 and $\sharp$3 which have the largest critical current. Since this effect complicates complicates the discussion, we will focus here on the 6K data. We note that sample$\sharp$1 has a critical current of the same order of magnitude than reported in Nb with very large grain using magneto optical imaging \cite{MO}. Compared to the sample $\sharp$1, samples $\sharp$2 and $\sharp$3 present a much larger critical current i$_c$ but their field variation looks very similar. For the whole magnetic field range B$<$B$_{c2}$, we find i$_{c\sharp 2}$/i$_{c\sharp 1}$ $\approx$ 4.8 and i$_{c\sharp 3}$/i$_{c\sharp 1}$ $\approx$ 5.6. The critical current of sample $\sharp$4 exhibits a field variation different from the others samples, with a strongest relative increase for B$\lesssim$ 0.125 T.  

Note also that the field dependence of $i_c$(B) changes at B$\approx$B$_{c2}$ for all the samples, signing the passage from the Shubnikov phase to the SSC phase. However, the change can be small or strong depending on the surfaces. In the following, we will discuss the main aspects of our observations.

\subsection{Critical current in the surface superconducting state}

Let us first focus on the SSC state, for B$_{c3}$ $>$ $B$ $>$ B$_{c2}$. The magnitude of $i_c$ changes impressively with the roughening of surfaces and it can be varied between less 0.5 A/cm up to 50 A/cm. We first note that 50 A/cm (J$_c$ $\approx$ 4.7 10$^3$ A/cm$^2$) is a significant value. It shows that the critical current can be important in the SSC state, notwithstanding the absence of superconducting volume. This reveals the strong ability of superconducting surfaces to transport non dissipative currents \cite{MS}. For the sample $\sharp$4 which has been polished using the colloidal silica, the SSC current is almost negligible.

The SSC sheath is defined by the distance over which the order parameter varies from the surface, and is of the order of 0.7 $\xi$ \cite{degennes}. Theoretically, the critical field of SSC B$_{c3}$ is maximum and close to 1.7 $\times$ B$_{c2}$  if it is oriented parallel to a boundary (B$_{\parallel }$), and tends to B$_{c2}$ in the opposite perpendicular geometry (B$_{\bot}$) \cite{yama}. However, this is true at the local scale, and it is likely that a real surface presents some asperity at the scale of $\xi$ along which SSC can nucleate. It was then proposed that report of SSC for B$_{\bot}$ could be due to the surface roughness \cite{patrice,niobsurfnoise}. In this case, the SSC critical currents could be reduced (increased) after making the perpendicular surfaces smoother (rougher), in agreement with our observations. This indicates also that the contribution of the (small) lateral faces to the SSC critical current is not important in our geometry. The SSC sheath is populated by small flux spots, known as "Kulik vortices", which are expected to be pinned and to move similarly as Abrikosov vortices do in the regular mixed state \cite{niobsurfnoise,Kulik}. It is likely that a significant part of the SSC critical current arises from a strong pinning of the small kulik vortices along the rough surface. We find also that the ratio B$_{c3}$/B$_{c2}$ increases slightly above the Saint James-De gennes value, reaching 1.84 for the rough samples $\sharp$2 and $\sharp$3. This is however much lower than the values reported in the case of near-surface interstitial contamination \cite{conta}.

\subsection{Critical current in the Shubnikov state}

Let us now discuss the Shubnikov mixed state, where vortices populate the bulk of the superconductor. It was already shown that the critical current of Niobium for B $<$ B$_{c2}$ can be also significantly modified by surface treatments. For example, after a surface irradiation using a low energy beam, the transport critical current of a Nb slab has been increased by a factor of 2.9 for a field range 0.83.B$_{c2}$ $<$ B $<$ B$_{c2}$) \cite{Niobiumnoise}. This increase was quantitatively accounted for by the increase of roughness in the surface topography. We observe here even a strongest increase of $i_c$ for samples $\sharp$2 and $\sharp$3 that extends over the whole magnetic range B $<$ B$_{c2}$. BCP treated samples such as the sample $\sharp$1 can present near surfaces impurities (e.g. hydrogen) that may act as pinning center. However, since the critical current increases after abrading this sample, we propose the surface roughness to be the major pinning source. To be more quantitative, we use the continuum approach proposed by Mathieu-Simon, well suited to describe the effect of rough surfaces on the vortex pinning \cite{MS}. The mean idea is that, due to the boundary condition on a normally rough surface, a lot of metastable states allows for the presence of an equilibrium, non dissipative current. At a macroscopic scale, this current can be written $i_c$$\approx$ $\varepsilon$ sin $< \theta>$ where $\varepsilon$ is the vortex potential ($M$=$\int_{Vol} -\varepsilon dV$ is the reversible magnetization, $V$ is the volume of the sample ) and $< \theta>$ is a mean contact angle. In a previous experiment, from a statistical analysis of surface roughness on a BCP Niobium such as sample $\sharp$1, a random surface disorder with mean angle of 2.2 deg was calculated. This angle explains a critical current $i_c \approx$ 10 A/cm close to B$_{c2}$  measured by V(I) curves \cite{Niobiumnoise}. Note first that the $i_c$ values that we calculate here from the magnetization loops are rather close, what gives confidence on the procedure of critical current extraction using the Bean model. For B$_{c2}$=0.195 T and a Ginzburg Landau parameter $\kappa$ $\approx$ 0.9 for pure Niobium and using the iteration procedure described by Brandt \cite{EHB}, we calculate the vortex potential for the whole magnetic field range (inset of fig.4). As shown in the fig.4, the critical current is then reproduced by the surface pinning approach described above with a critical angle $\theta$ $\approx$ 5.2 deg. To explain an increase of $i_c$ by a factor of 4.8 (resp. 5.6) between the samples $\sharp$1 and $\sharp$2 (resp.$\sharp$3), the critical angle should reach values of around 26 deg (resp. 31 deg). Those are large values but they still have a physical meaning. At first sight, it could be surprising that sample $\sharp$3 presents a larger critical current than sample$\sharp$2, whereas 2400 sandpaper is rougher than 3 $\mu$m diamond. However, this contradiction is only apparent since more than the rms roughness, the important parameter for surface pinning is the local slope of the surface defects. We find also that further roughnening has only marginal effect on the critical current of sample $\sharp$3, and it is tempting to conclude that a critical angle close to 45 deg is a sort of maximum geometrical angle. The surface critical current of sample $\sharp$3 is likely the maximum that can be reached in Niobium by introducing surface defects, and it is close to 600 A/cm at low field. 

\subsection{The special effect of colloidal silica}

Colloidal silica is currently used to polish wafers to low-defect and ultra-flat surfaces. It is specially efficient to smooth out most of the surface bumps.  
The sample $\sharp$4 which has been polished with colloidal silica (50 nm monodisperse particles with size distribution of roughly 50 \% ) shows two peculiar effects on its critical current. As discussed above, the SSC critical current is extremely small in this sample. This indicates that the \textit{planarization} of the surface has been extremely efficient and that the surface bumps have been strongly reduced at the scale of 0.77 $\xi$ $\approx$ 30 nm, what is indeed close to the size of silica particles. To give more support to this view, we have measured the critical current for the two geometry: B$_{perp}$ and  B$_{para}$ (Fig.5). As discussed before, only the large facets have been polished. The SSC critical current strongly collapses for the B$_{perp}$ geometry as expected for planar perpendicular surfaces, whereas it is still measurable for the B$_{para}$ geometry. The sample $\sharp$4 shows the conventional SSC behavior observed for an ideal smooth surface parallel to the magnetic field.
In addition, the critical current has a non monotonic variation as function of the magnetic field for B $<$ B$_{C2}$. It is slightly smaller than the reference sample $\sharp$1 for B close to B$_{c2}$ but is significantly larger for B $<$ 0.125 T. This means that, compared to the sample $\sharp$1, the pinning potential has been reinforced when it is probed by vortex with spacing larger than $(\phi_0/B)^{1/2} \approx$ 130 nm. One possibility is that the surface roughness and/or corrugation are not random for all the spatial scale and present some correlations. Since we are here interested in large area views of the variations in surface structure, and in order to investigate this effect, we have imaged the samples surfaces using scanning electron microsocopy (in the secondary electrons collection mode). In the following, we will assume that the colors (256 gray scale using Tagged Image File Format) corresponds to an effective height of surface, what is in principle expected for a chemically homogeneous surface. However, in absence of rigorous calibration, it is not possible to extract absolute values in the third direction. We will then limit the discussion to a qualitative comparison of the topography and corrugation between the samples. Different SEM images recorded under similar experimental conditions are shown in the Fig.6.  A strong difference in surface topography can be easily noted, as expected from the different surface treatments. Note that the images are representative of the major part of the samples in the sense that we have observed very similar image and statistics over different surface areas. Sample $\sharp$2 and $\sharp$3 exhibit damaged surfaces and have the the largest critical current, in agreement with reinforced vortex pinning by surface defects. Note that some Niobium hybrides can be observed on the sample $\sharp$4 \cite{hybrides}, but they are very rare (one or two in a 100 $\times$ 100 $\mu$m$^2$ area) and have likely no effect for vortex pinning. To reveal if any other peculiarities exist the sample $\sharp$4, we have studied the difference in surface topographies using the power spectral density (PSD), as proposed in \cite{Niobiumnous, Xu}. In details, we have numerically calculated the autocorrelation function C(r,r$_0$)=$<$h(r+r$_0$)h(r$_0$)$>$ from the greyscale images (h(r)) of samples $\sharp$3 and $\sharp$4. From the Wiener-Khinchin theorem, the PSD can be deduced from the Fourier transform of this autocorrelation function.
\begin{equation}
S_q=1/2\pi^2 \int C(r,r_0)  e^{-iq.r} d^2r 
\end{equation}
q is the spatial frequency and is in inverse-length units.
We recall that a random, uncorrelated, surface topography would give a flat PSD (white noise), whereas a q-dependence of the PSD indicates some correlation in the surface topography. The PSD is specially adapted to reveal periodic or quasi-periodic surface features. In general, PSD of surfaces exhibit a power law dependence S$_q$ $\propto$
 q$^{-\alpha}$ \cite{Xu}.
As shown in fig.7, the SEM image of sample $\sharp$3 presents a typical q$^{-\alpha}$ PSD with $\alpha \approx$ 1.8, not far from a Lorentzian distribution of surface heights 
($\alpha$=2). For the sample $\sharp$4, the PSD is notably different (fig.8). In addition to the q$^{-\alpha}$ variation with $\alpha \approx$1.6, different broad peak can be observed for q$<$0.007 nm$^{-1}$. Note that they are not simple harmonics of the largest one since they are not at commensurate position. They indicate different quasi-periodicity in the surface topography, for a spatial range larger than d$^{*} \approx$1/0.007$\approx$143 nm. It is reasonable to deduce that the surface pinning potential will be also more efficient for vortex separated by more than d$^{*}$. Remarkably, this vortex spacing corresponds to a magnetic field lower than $\phi_0/d^{*2}\approx$0.1T, close to the field where the critical current of sample $\sharp$4 appears to be reinforced. We conclude that the effect of colloidal silica is likely double, it smooths out surface bumps and strongly decreases the SSC for the perpendicular field geometry, but also induces a peculiar corrugation at low spatial scale that acts as efficient pinning center for (Abrikosov) vortex lines in the Shubnikov state.

\section{conclusion}
To conclude, we have observed that relatively simple surface treatments lead to important change of the critical current in Niobium slabs, both in the Shubnikov and in the surface superconducting states. In particular, the critical current in the surface superconducting state can be increased or decreased by more than two orders of magnitude, emphasizing the strong ability of superconductor surfaces to transport non dissipative currents. A qualitative interpretation of the results has been proposed, and a more precise analysis of the surface roughness and topology using AFM would be helpful to go further insight microscopic pinning mechanisms. Of course, it would be also interesting to test some other surface treatments at different spatial scale and to analyze consequences on the pinning and transport properties. We note also that a proper choice of surface treatments and a detailed analysis of the changes in macroscopic properties can be very helpful to separate bulk superconductivity and surface superconductivity with bulk normal conduction, a problematic of renewest interest due to the emergence of topological superconductors.

Acknowledgments: A.P thanks the program ECOS-MINCyt A14E02 for support.

\newpage

\begin{figure}[p]
\begin{center}
\includegraphics*[width=10.0cm]{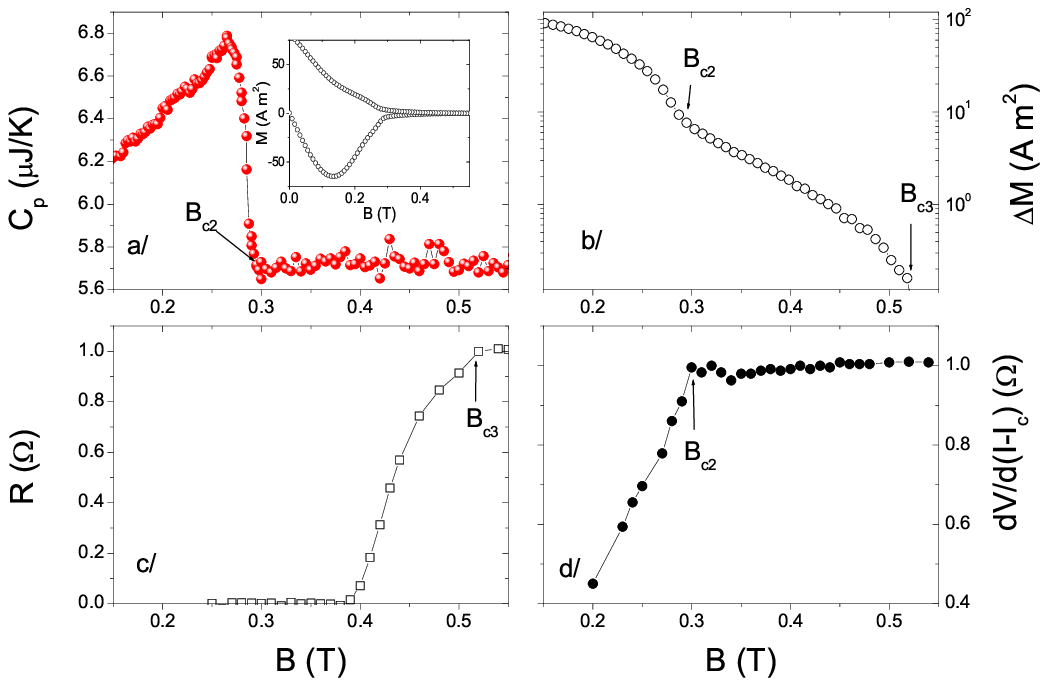}
\end{center}
\caption{Characterization of a Niobium slab after chemical polishing. All measurements are done at T=4 K. a/ Specific heat as function of the magnetic field. In the inset is shown the magnetic moment loop as function of the field. b/ Irreversible part of the magnetic moment as function of the magnetic field. c/ Electrical resistance as function of the magnetic field with I=0.1 A (i$\approx$ 2 10$^{-4}$ A/m). Since the biased current is lower than the critical current in the surface superconducting state, the resistance is still zero for B$>$ B$_{c2}$. d/ From V(I) curves measured at different field values, we  report the differential resistance R$_i$=dV/d(I-I$_c$) as function of the field. In the surface superconducting state, V=R$_n$(I-I$_c$) and R$_i$=R$_n$ precisely for B=B$_{c2}$.} 
\label{fig.1}
\end{figure}

\begin{figure}[t!]
\begin{center}
\includegraphics*[width=10.0cm]{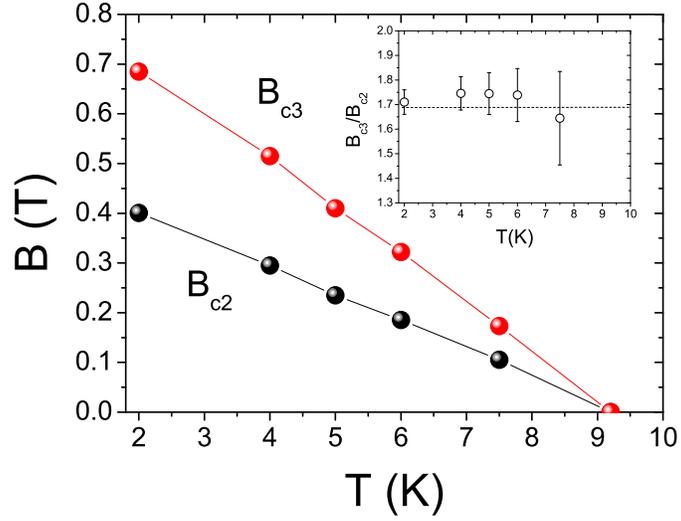}
\end{center}
\caption{Phase diagram (B,T) of a Niobium slab after chemical polishing (BCP). In the inset is shown the ratio B$_{c3}$/B$_{c2}$. The dotted line is the Saint James-De Gennes ratio 1.69.}  
\label{fig.2}
\end{figure}

\begin{figure}[t!]
\begin{center}
\includegraphics*[width=8.0cm]{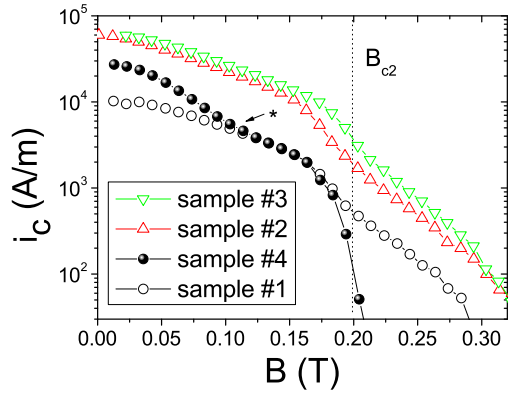}
\end{center}
\caption{Critical current $i_c$=J$_c$ $\times$t/2 of the samples $\sharp$1 (BCP), $\sharp$2 (2400 sandpaper), $\sharp$3 (3 $\mu$m diamond) and $\sharp$4 (colloidal silica) at T= 6K. The stars indicates the field B$\approx$ 0.125 T where the critical current changes its relative variation for sample $\sharp$4. The magnetic field is perpendicular to the large facets.}  
\label{fig.3}
\end{figure}

\begin{figure}[t!]
\begin{center}
\includegraphics*[width=8.0cm]{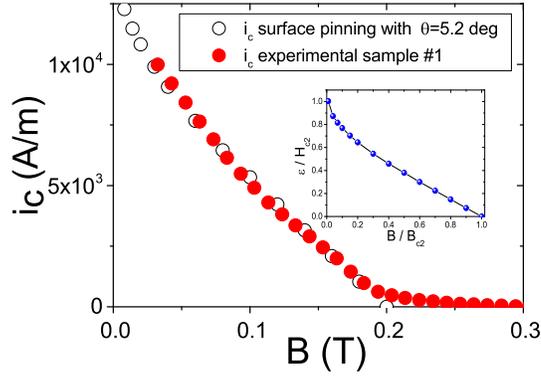}
\end{center}
\caption{Critical current $i_c$=J$_c$ $\times$t/2 of the sample $\sharp$1 at T= 6K. Also shown is the theoretical surface critical current using a surface pinning model with a critical angle $\theta$=5.2 deg and a potential $\epsilon$ shown in the inset (see text for details).}  
\label{fig.4}
\end{figure}

\begin{figure}[t!]
\begin{center}
\includegraphics*[width=10.0cm]{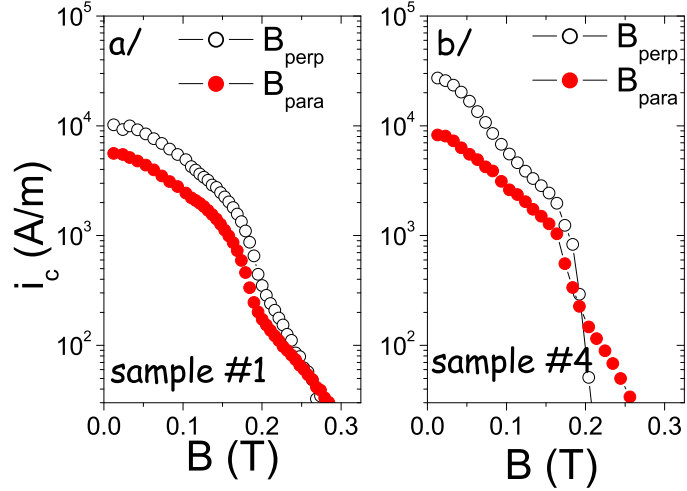}
\end{center}
\caption{The surface critical current of the samples $\sharp$1 (BCP) (a/)and $\sharp$4 (colloidal silica) (/b) at T= 6K, for the two orientations of the magnetic field (B$_{perp}$ and B$_{para}$, for B respectively perpendicular and parallel to the large facets). Note the collapse of surface currents for B$>$Bc2 for the sample $\sharp$4 in the B$_{perp}$ geometry.}  
\label{fig.5}
\end{figure}

\begin{figure}[t!]
\begin{center}
\includegraphics*[width=10.0cm]{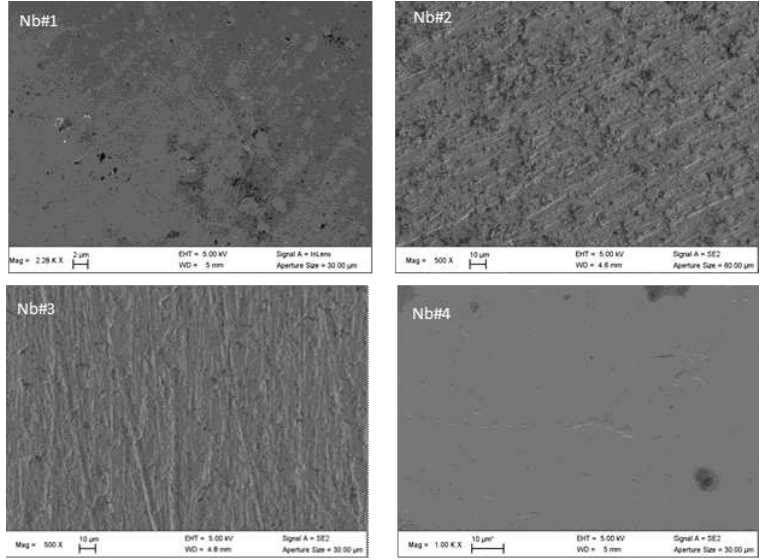}
\end{center}
\caption{SEM images of the surfaces of the Niobium samples after different surface treatments (sample $\sharp$1 (BCP), $\sharp$2 (2400 sandpaper), $\sharp$3 (3 $\mu$m diamond) and $\sharp$4 (colloidal silica).}  
\label{fig.6}
\end{figure}

\begin{figure}[t!]
\begin{center}
\includegraphics*[width=10.0cm]{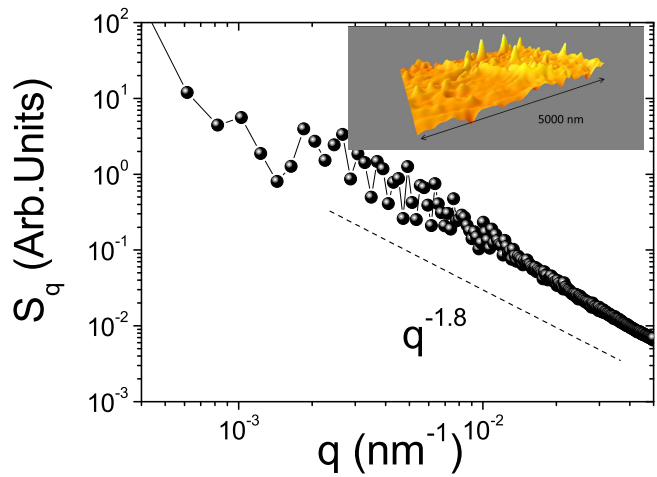}
\end{center}
\caption{Power spectral density of the SEM image of sample $\sharp$3 (3 $\mu$m diamond). In the inset is shown a 3D view of the SEM image.}  
\label{fig.7}
\end{figure}

\begin{figure}[t!]
\begin{center}
\includegraphics*[width=10.0cm]{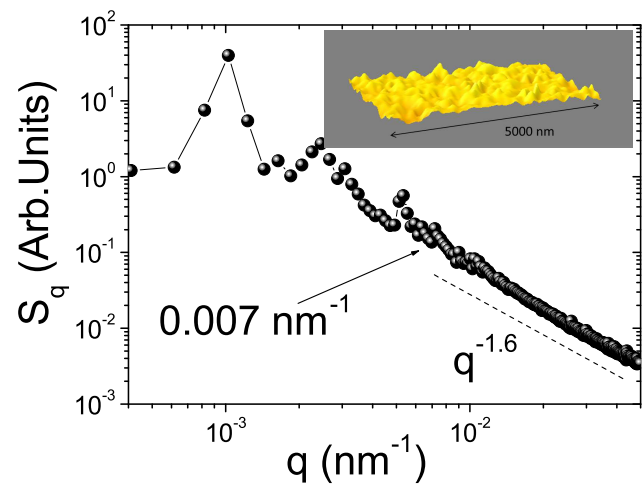}
\end{center}
\caption{Power spectral density of the SEM image of sample $\sharp$4 (colloidal silica). Note the different peaks at low q, appearing for q$<$0.007 nm$^{-1}$. In the inset is shown a 3D view of the SEM image.}  
\label{fig.8}
\end{figure}

\end{document}